\documentclass[11pt, a4paper, copyright, gr]{googledan}

\usepackage[numbers, sort&compress]{natbib}
\usepackage{longtable}
\usepackage{supertabular,booktabs}
\usepackage{float}
\usepackage[capitalize,noabbrev]{cleveref}

\makeatletter
\let\oldlt\longtable
\let\endoldlt\endlongtable
\def\longtable{\@ifnextchar[\longtable@i \longtable@ii}
\def\longtable@i[#1]{\begin{figure}[t]
\onecolumn
\begin{minipage}{0.5\textwidth}
\oldlt[#1]
}
\def\longtable@ii{\begin{figure}[t]
\onecolumn
\begin{minipage}{0.5\textwidth}
\oldlt
}
\def\endlongtable{\endoldlt
\end{minipage}
\twocolumn
\end{figure}}
\makeatother

\uselogo{} 

\title{Adaptive Cardio Load Targets for Improving Fitness and Performance}

\correspondingauthor{justinphillips@google.com}

\renewcommand{\today}{2025-08-15}

\author[1]{Justin Phillips}
\author[1]{Daniel Roggen}
\author[1,2]{Cathy Speed}
\author[1]{Robert Harle}

\affil[1]{\thepa{}{}}
\affil[2]{Work done while at Google}

\begin{abstract}
{\em Cardio Load}, introduced by Google in 2024, is a measure of cardiovascular work (also known as training load) resulting from all the user's activities across the day. It is based on heart rate reserve and captures both activity intensity and duration. Thanks to feedback from users and internal research, we introduce adaptive and personalized targets which will be set weekly. 
This feature will be available in the Public Preview of the Fitbit app after September 2025. 
This white paper provides a comprehensive overview of Cardio Load (CL) and how weekly CL targets are established, with examples shown to illustrate the effect of varying CL on the weekly target. We compare Cardio Load and Active Zone Minutes (AZMs), highlighting their distinct purposes, i.e. AZMs for health guidelines and CL for performance measurement. We highlight that CL is accumulated both during active workouts and incidental daily activities, so users are able top-up their CL score with small bouts of activity across the day.

\end{abstract}

\begin{document}

\maketitle

\section{Introduction to Cardio Load}
\subsection{Rationale for Cardio Load measurement}

Engaging in structured physical activity promotes positive physiological, physical, and psychological changes, which in turn support the maintenance or improvement of overall health and fitness \cite{warburton2006}. 
Global health guidelines on exercise intensity and its relationship with long-term health outcomes, particularly mortality, are well-established through extensive research \cite{piercy2018}. 
Due to a lack of precise intensity measurements, exercise intensity is often broadly categorized as `light,' `moderate', or `vigorous', however an increasing body of research supports the importance of the full spectrum of physical activity, from gentle to intense  \cite{saintmaurice2018,fuzeki2017}.

There is a dose-response relationship of exercise such that, while higher-intensity exercise may demonstrably yield greater and broader health benefits such as reduced mortality, light activities also have a positive impact, in particular on cardiometabolic health. For those new to exercise who may be intimidated by higher intensities or who have physical limitations, starting with light intensities can be effective. Moreover, even for those engaging in regular vigorous exercise, incorporating light activity into their routine can also aid recovery \cite{greco2012, west2013}. 

The dose-response relationship also implies that high intensity physical activity such as interval training can have particular health and performance benefits \cite{shiraev2012}, so users should be able to track and gain appropriate reward for incremental intensity levels beyond the moderate-to-vigorous threshold. Google introduced {\em Cardio Load} in 2024 \cite{Speed24b} to provide users of its fitness ecosystem with such a quantitative reward.

\subsection{The Cardio Load calculation}

Cardio Load (CL) is based on the concept of ``TRaining IMPulse" (TRIMP), which defines cardiovascular load as the product of the duration of an exercise session with its intensity \cite{bannister1991, morton1990}. 
Earlier TRIMP models used average percentage heart rate reserve ($\%HRR$) measured across a specific workout. 
However, the proliferation of heart rate monitors now allows continuous and more granular estimation of intensity at the minute level.
One of the simplest calculations of intensity is the $\%HRR$ itself, but Cardio Load adopts and extends a more sophisticated TRIMP model from Banister \cite{bannister1991} that exponentially weights the intensity value as $\%HRR$ increases:

\begin{equation}
L_{Banister} = 0.64 \cdot \%HRR/100 \cdot e^{k \cdot \%HRR/100}
\end{equation}

\begin{equation}
\text{with }\%HRR = \frac{(HR - RHR)}{(HR_{max} - RHR)} \times 100\%%
\end{equation}

Where $HR$ is the average heart rate during one minute, $RHR$ is the individual's resting heart rate, $HR_{max}$ is the maximum heart rate, $e$ is the exponential function and $k$ is a weighting factor of 1.92 (male) or 1.67 (female), the difference being accounted for by the fact that average heart rates are higher for females than males for a given physiological load \cite{rascon2020}. 

The exponential recognises that the strain imposed by training is not a continuum, in other words higher sustained intensities impose disproportionately higher strain. Furthermore, high intensity interval type workouts impose different strain than less intensive steady state work \cite{gibala2012}. We therefore have a preference for the exponential formulation rather than directly applying $\%HRR$ \cite{kellmann2010,macinnis2017}. 

Cardio Load also extends Banister's original formulation in three ways to ensure its relevance and accuracy in 24/7 tracking:
\begin{itemize}
\item
A minimum of 30\% $HRR$ (the low end of the light intensity zone) is needed to obtain load. 
This recognises that the original formulation did not anticipate gathering heart rate data outside of a structured training session, when low heart rates are dominant. 
A non-zero threshold is chosen because very light physical activities are insufficient for comprehensive systemic health improvements \cite{blair1996}.
\item
Each minute accruing load must have evidence of movement, which is done through analysis of the inertial sensors on the smartwatch or tracker. 
This reflects that the heart rate can be elevated through psychological stress rather than physical exertion, but this has been associated with {\em negative} cardiovascular outcomes \cite{chida2010} rather than health or performance improvements, so is discarded.
\item
We down-weight loads between 30\% and 40\% $HRR$. 
This recognises that the current evidence supports moderate-to-vigorous physical activity (MVPA) having a disproportionately greater impact on health than lighter work \cite{chastin2019,kim2022,ku2020,qiu2021}.
\end{itemize}

In addition all auto-detected workouts and workouts manually logged by the user contribute to Cardio Load.

Cardio Load is thus a metric that represents the load imposed on the cardiovascular system across the day, ranging from light work to peak effort, and recognises that higher intensities result in exponentially greater stress on the system \cite{kellmann2010,macinnis2017}. 
Notably, even resistance training, which focuses primarily on muscular load, imposes some cardiovascular load that Cardio Load will capture. 

The graph in \cref{fig:cl_minute} below depicts the accrual of Cardio Load as a function of heart rate reserve. The `steps' in the graph occur at the threshold of where Cardio Load is first awarded ($\%HRR$ above 30\%); and where the downweighting ends ($\%HRR$ of 40\%). 

\begin{figure}[hbt]
\centering
\includegraphics[width=.8\linewidth]{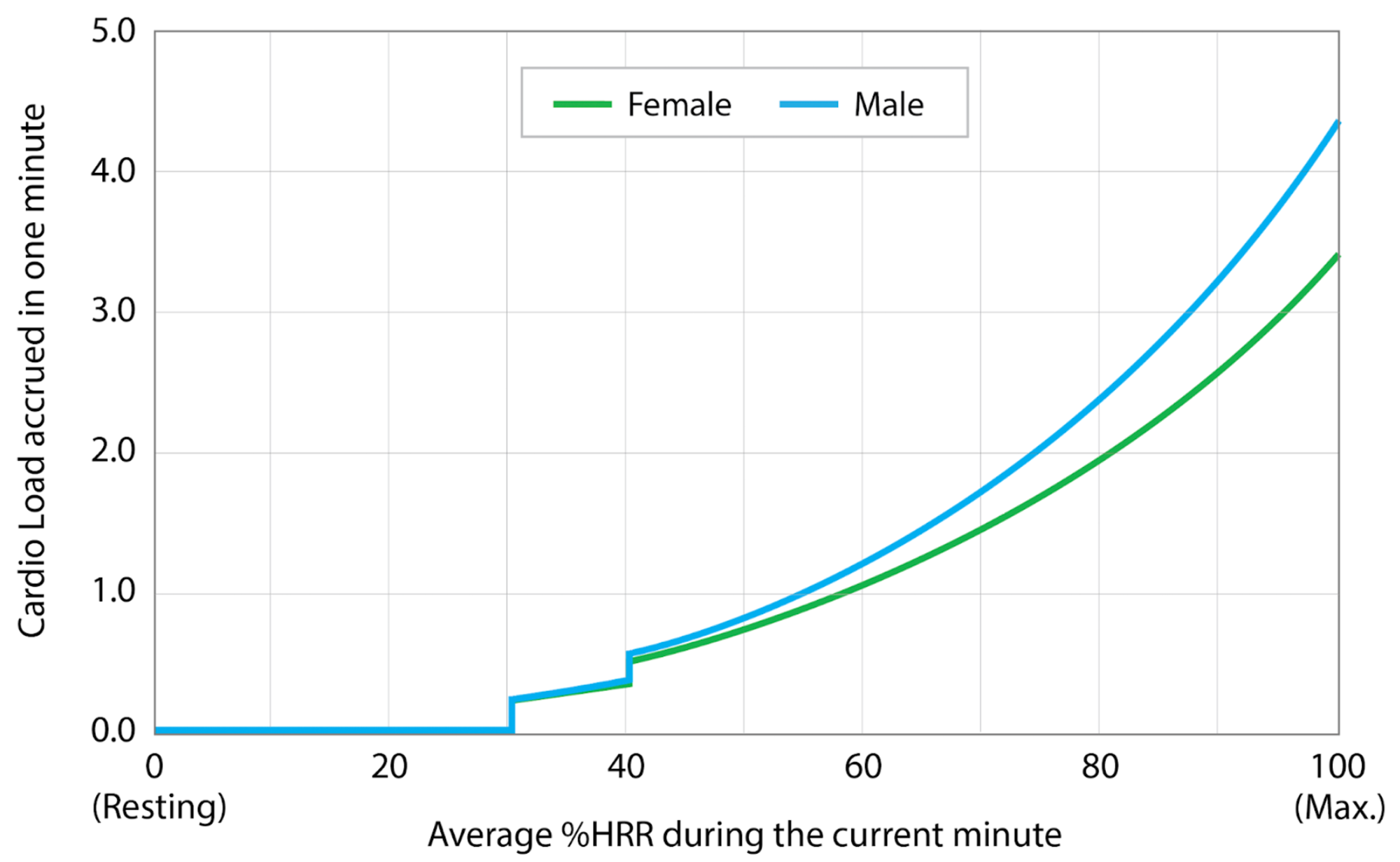}
\caption{Graph of Cardio Load accrued in a minute vs the \%Heart Rate Reserve during a minute.}
\label{fig:cl_minute}
\end{figure}

\subsection{Why go beyond Active Zone Minutes?}

Cardio Load is designed to complement an earlier load metric known as Active Zone Minutes (AZMs) that was developed to reflect global physical activity guidelines. Although similar, there are fundamental differences between CL and AZMs:

{\em Adherence to health guidelines.} 
AZMs are designed explicitly to represent the health guidelines. 
Recommendations \cite{piercy2018} (and hence AZMs) are focused on population guidelines for increasing longevity and health, rather than on performance measurement and improvement. 
Although AZMs are ostensibly simple, they can be difficult to interpret without clarifying materials. 
For example, AZMs derive from time spent in the moderate and vigorous intensity zones, but user research has shown that users are often unsure whether they are exercising in either zone, or even what these definitions mean in practice.

{\em Improvement of performance.} CL is based on an established way to measure load for the purposes of performance. 
While AZMs can also be framed as a form of TRIMP where intensity is bucketed into just three states (light (0 AZM), moderate (1 AZM), and vigorous (2 AZMs)) and are well publicized in health guidelines,  CL provides a more precise measurement that both extends to higher intensities and to lower intensities. 
As such, CL and AZMs are complementary, but CL will reward small bouts of intense activity beyond the vigorous threshold, and therefore will recognise the incremental changes that are critical to enhancing performance. Furthermore, CL being more sensitive to shorter bursts of cardiovascular load will also proportionately reward resistance training.

\section{Acquiring Cardio Load during workouts and daily activity}

\begin{figure}[hbt]
\centering
\includegraphics[width=1\linewidth]{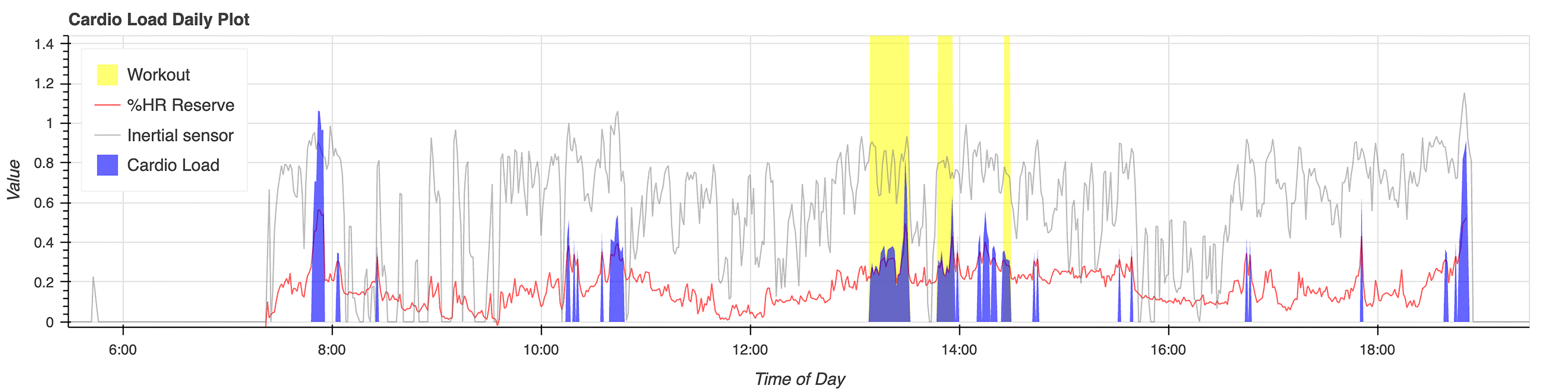}
\caption{Example of a day's activity showing how Cardio Load accumulates across the day during workouts and incidental daily activity.}
\label{fig:cl_day}
\end{figure}

Users accumulate Cardio Load during workouts as well as when engaging in incidental daily activity that sufficiently raises the heart rate such as short brisk walks, commuting or climbing stairs. 
We will refer to CL accrued through workouts (both manually tracked and auto-detected) as `workout load' and CL from incidental daily activities as `incidental load'. 
\cref{fig:cl_day} shows an example of a day's activity showing how a user accumulates CL. In this case, the user gained a CL of 37 by the end of the day. CL increments are shown in blue together with $\%HRR$ (red) and motion as indicated by the inertial sensors (grey). Workout load is highlighted in yellow, while some incidental load (approximately 45\% of the total daily CL) was accumulated at various times across the day.

\begin{figure}[hbt]
\centering
\includegraphics[width=.8\linewidth]{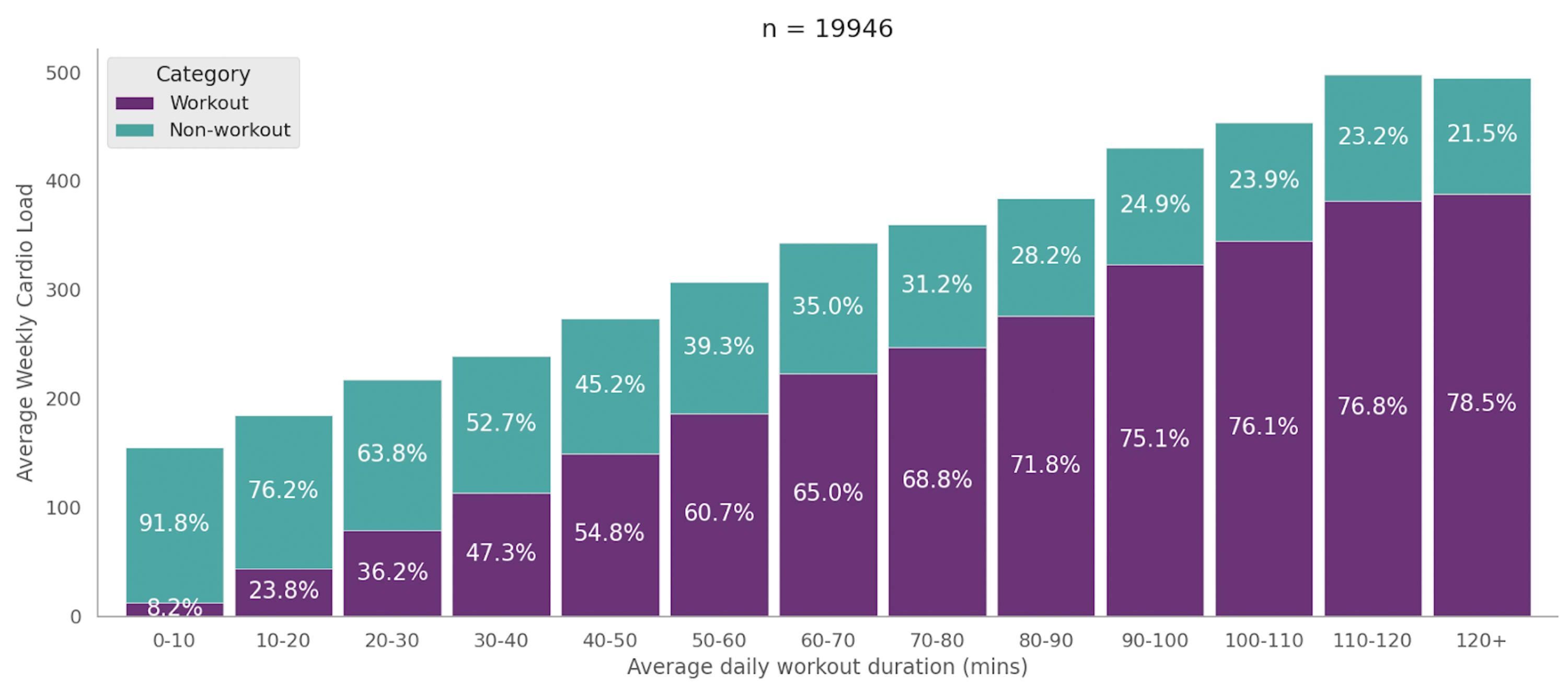}
\caption{Weekly Cardio Load and proportion of CL that is allocated during workout and daily activity for different ranges of mean daily workout duration.}
\label{fig:cl_week}
\end{figure}

Users with regularly logged workouts will acquire a significant proportion of their Cardio Load during this time, and can further top-up their CL score during daily activity. 
The graph in \cref{fig:cl_week} shows the average weekly CL for all users for different ranges of mean daily workout duration. The graph also shows the proportion of CL that is allocated during defined workouts and incidental daily activity.

As we would expect, total CL increases with increasing time spent engaging in workouts, as does the proportion of CL earned while working out. In the group of twenty thousand Fitbit users studied, those whose mean workout time exceeded two hours still accumulated at least 20\% of their CL on average during daily activity. 
Newcomers to the fitness journey will also benefit from CL. A user who only completes an average of 10 minutes workout per day will still accumulate around 180 CL on average, with around 75\% coming from daily activity.

\begin{figure}[hbt]
\centering
\includegraphics[width=.8\linewidth]{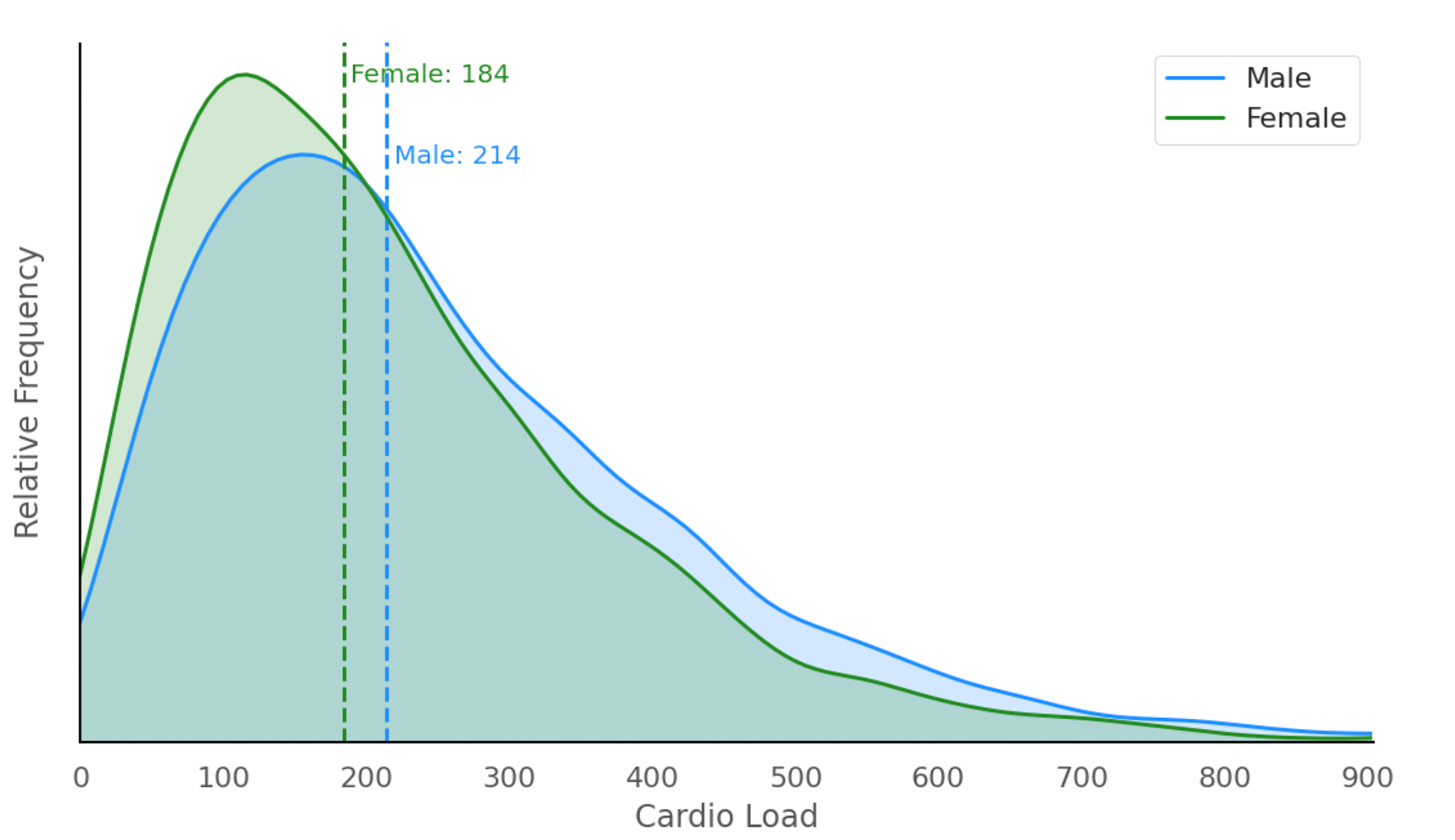}
\caption{Histogram of weekly Cardio Load in Fitbit users averaged over several weeks (n=19,946).}
\label{fig:cl_hist}
\end{figure}

\cref{fig:cl_hist} shows a histogram of weekly CL in twenty thousand consenting Fitbit users averaged over several weeks. The median weekly CL is 214 for males and 184 for females.

\section{Cardio Load Targets}

The Cardio Load Target we set for a user is intended to guide them to maintain or increase their cardiorespiratory fitness while also providing a mechanism to alert them wheret excessive increases or decreases in load are detected (neither of which is generally good, with both increasing the chance of injury). 
In the literature Acute:Chronic Workload Ratio (ACWR) is the main mechanism that is used to assess and contain the load.  
An ACWR of approximately 1 is intended to maintain the user’s internal load and, by proxy, their fitness level. 
ACWR values outside of a narrow range around 1 are considered problematic as they imply that the user is at risk of either reducing their fitness level or over-training.

Approach: 
\begin{itemize}
\item
The Cardio Load target is set weekly, at a level that represents what a user should achieve across the week to maintain their cardiorespiratory fitness (i.e. equivalent to the chronic load or an ACWR of approximately 1).
\item
We encourage users to exceed the CL Target for increased fitness, and also warn them if we detect that they overdo it.
\item
We set a minimum CL Target value, representing something everyone should get, regardless of how (in)active they are. 
\end{itemize}

Setting target load is critical for motivation and optimal training effectiveness. In order to avoid overtraining the target should not be set too high, while a target that is too low would miss optimal training opportunities.

The default behavior of the target load is to guide the user to a level of activity that maintains their current fitness level. The approach we have chosen presents users with a target based on CL history that represents the user's `norm'. 
Of course there will be times that the target is unobtainable due to injury or busy schedules and other times where it feels too low. The target is designed to allow users to apply their judgement and aim higher or lower depending on their ambition to get fit balanced with how well they feel, their Readiness score or other factors. 

\subsection{From Daily to Weekly Cardio Load Targets}

\begin{figure}[hbt]
\centering
\includegraphics[width=.8\linewidth]{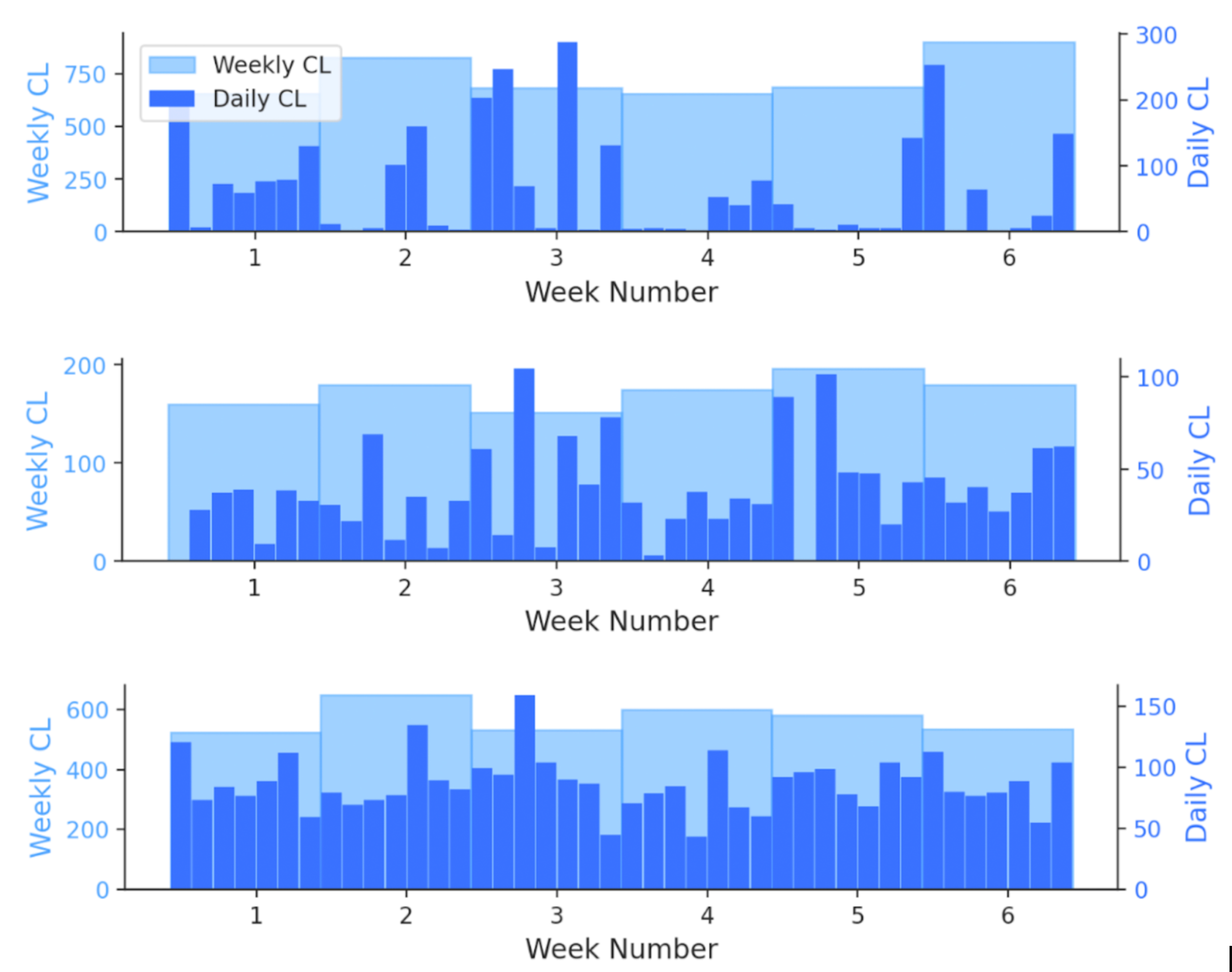}
\caption{Six weeks of daily (dark blue) and total weekly (light blue) Cardio Load for three typical users.}
\label{fig:cl_six_weeks}
\end{figure}

Many users concentrate structured physical activity on certain days only, often alternating higher training load days with days of light training or rest days. 
\cref{fig:cl_six_weeks} shows six weeks of daily (dark blue) and total weekly (light blue) Cardio Load for three typical users. It can be seen that the daily values fluctuate considerably, but this fluctuation is smoothed out by the total weekly CL.

Furthermore, traditional fitness coaching often assumes a weekly period, as this is in line with the cultural pattern of our lives. Coaches may seek to define a series of sessions that would be expected to provide loads in line with the `just right’ range, allowing for alternate periods of intense and lighter training or rest days.

\begin{figure}[hbt]
\centering
\includegraphics[width=.8\linewidth]{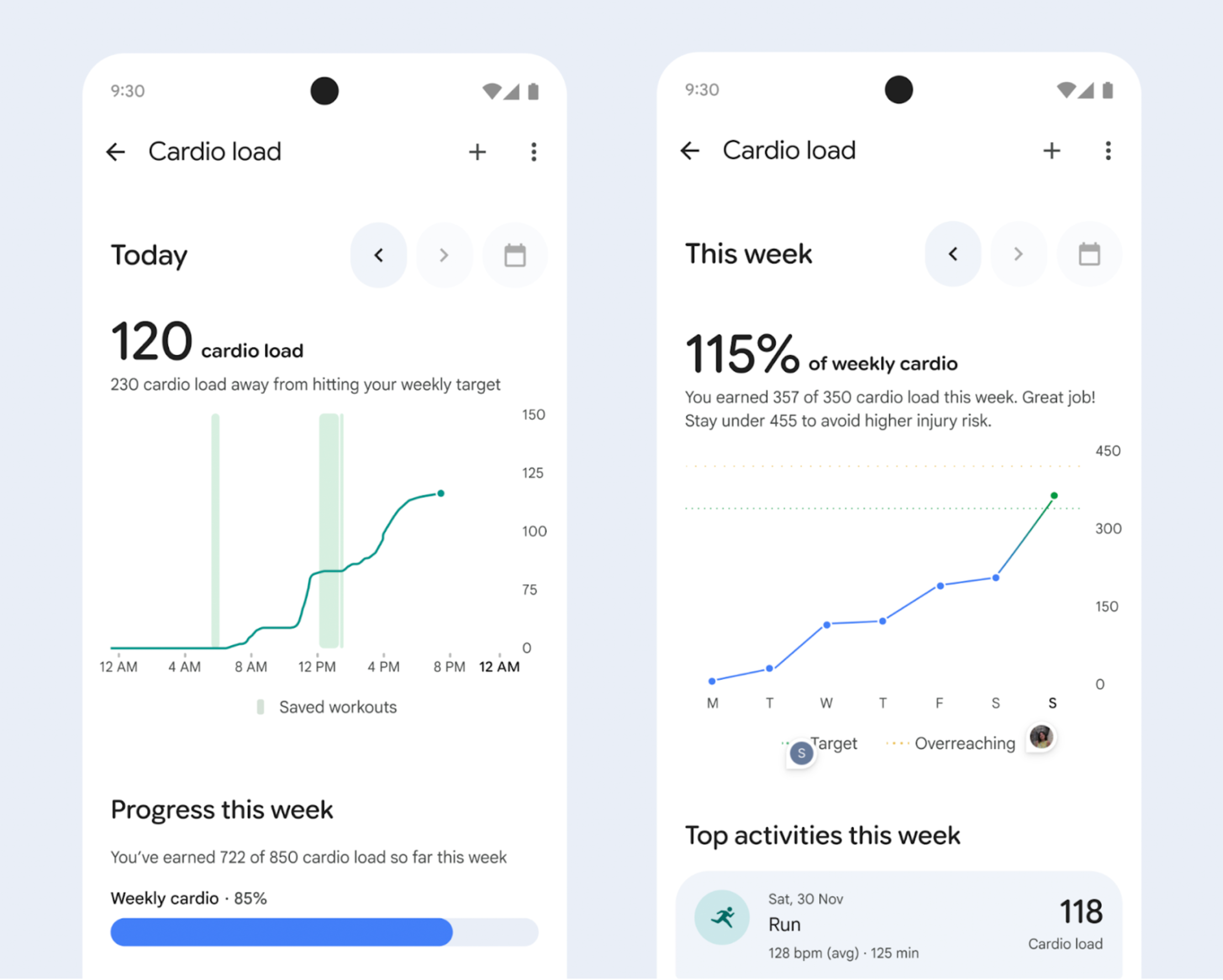}
\caption{Cardio Load daily (left) and weekly (right) views showing progress towards the weekly target load. Note: this is for illustration purposes; final appearance may vary.}
\label{fig:ui}
\end{figure}

While the Pixel Watch 3 Target Load was computed daily, internal research and customer feedback prompted a new approach. 
With the upcoming Public Preview of the Fitbit app, we will introduce a weekly CL Target. The daily accrued CL is still presented to the user (see \cref{fig:ui}) but our interaction design encourages the user to see this as an increment towards their weekly target.

In some cases users may find that they have a slow start to the week, for example due to illness or busy schedules, that will inevitably lead to falling behind the target. 
The weekly Cardio Load view enables the user to try to plan to catch up as the week progresses. 
Although it's not always possible to reach the target every week, this weekly perspective empowers users to take control of their fitness journey, allowing for flexible planning and consistent progress towards their goals.

\begin{figure}[hbt]
\centering
\includegraphics[width=.8\linewidth]{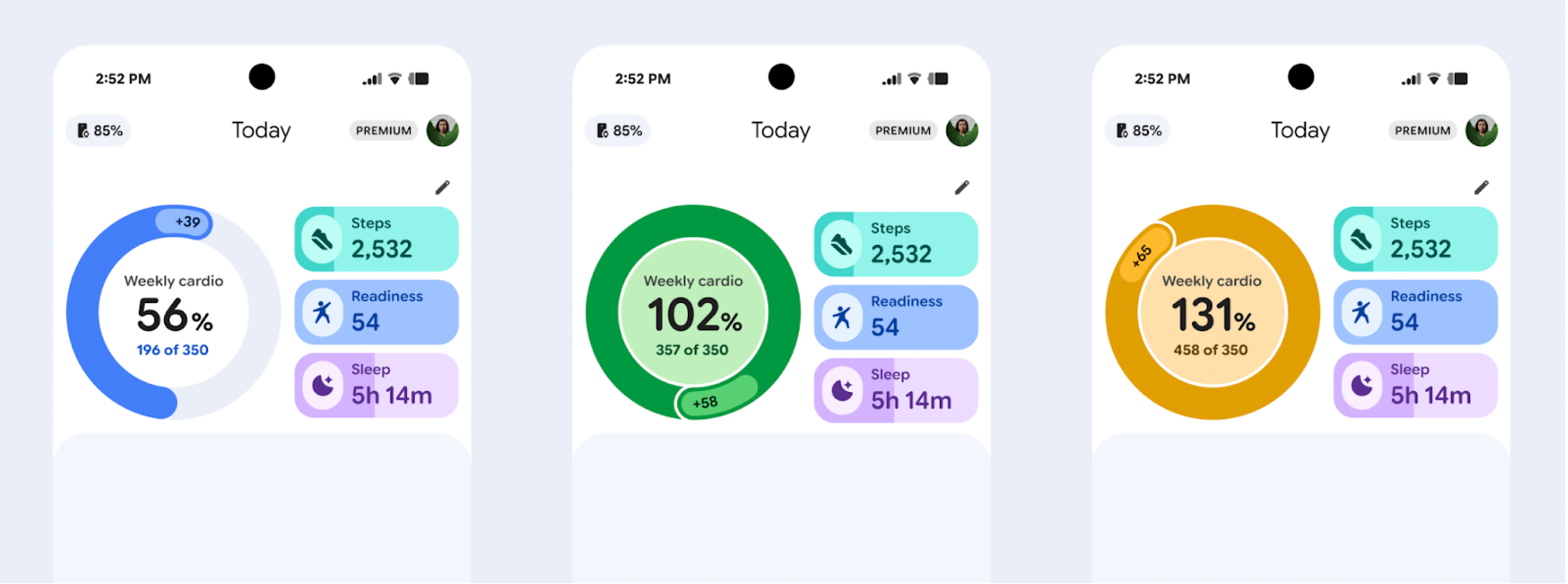}
\caption{High level glanceable view of Cardio Load in the Fitbit app.  Note: this is for illustration purposes; final appearance may vary.}
\label{fig:ui2}
\end{figure}

\cref{fig:ui2} shows the Focus Metrics view of Cardio Load in the Fitbit app. 
The ring indicates progress towards the weekly target, while the pale colored sector shows the Cardio Load increment for the current day. The blue ring in the left pane of the figure shows Cardio Load before the target is met. Once the target is met the ring turns green. If the user greatly exceeds the target the ring color changes as a warning of potential for over-training.

\subsection{How the Pixel Watch calculates an adaptive CL Target}

Human fitness coaches use history and knowledge to predict broadly what the optimum load might be. As a trainee's fitness changes, they are able evaluate the progress to the target load, and adapt the target accordingly. The CL Target adapts to users' progress in a similar way.

By definition, the chronic load referred to in ACWR is the typical load for that person in recent times. Setting a target in this model then requires us to estimate the chronic load. There are two established methods to compute the chronic load \cite{williams2017}:

\begin{itemize}
\item
{\bf Rolling mean (RM).} 
The mean weekly load over the previous 28 days (i.e. the mean of four weekly values).

\item
{\bf Exponentially Weighted Moving Average (EWMA).} 
This is an exponential weighting computed as $EWMA = \alpha \cdot CL(t) + (1 - \alpha) \cdot  CL(t-1)$ for some constant, $\alpha$ where $CL(t)$ is the latest 7-day Cardio Load and $CL(t - 1)$ is the previous week's 7-day Cardio Load.
\end{itemize}

Both of these are in wide use, although the constant a for EWMA varies across implementations. 
The chronic load can act as a proxy for fitness level (since increasing loads on the body elicit a fitness response). 
The EWMA response to short term change in weekly CL is dependent on the choice of the constant, $\alpha$; the smaller the constant, the less sensitive to change. 

\begin{figure}[hbt]
\centering
\includegraphics[width=.8\linewidth]{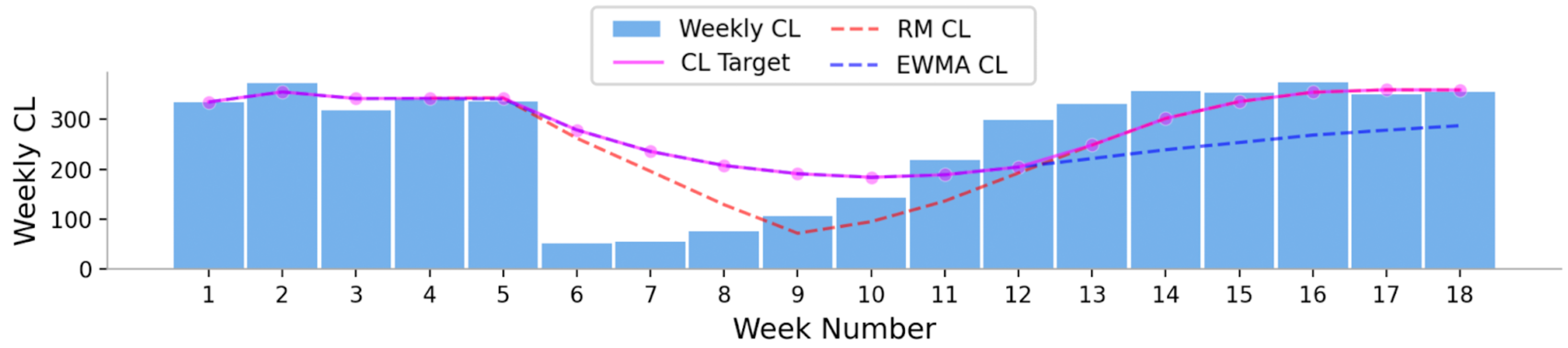}
\caption{Plot of Weekly CL, RM, EWMA and CL Target with a step decrease in weekly CL at week 6 followed by a gradual recovery.}
\label{fig:target_adaptation_1}
\end{figure}

\cref{fig:target_adaptation_1} shows a graph of weekly Cardio Load together with EWMA and RM from one user over an 18 week period. In this example the user's normally high Cardio Load is interrupted at week 6, perhaps due to illness or injury, and then gradually recovers. The RM (red dashed line) drops rapidly, while the The EWMA (blue dash) does not over-react to the break and keeps the target initially in the same ball park then falls gradually. However, when the user returns to their usual routine, it is very slow to rise. Even after 6 weeks of consistently achieving the same higher load, EWMA remains stubbornly below that load. By contrast, the rolling mean line (red dash) is much faster to return to the original when the user demonstrates that they can still achieve it. To get the best of both worlds, the CL Target (magenta) takes the maximum of the rolling mean and EWMA outputs.

Analysis of a large number of examples showed good performance for all scenarios when the maximum of RM and EWMA was used as the target, i.e.,

\begin{equation}
    CL_{Target}  = Max[RM, EWMA].
\end{equation}

\begin{figure}[hbt]
\centering
\includegraphics[width=.8\linewidth]{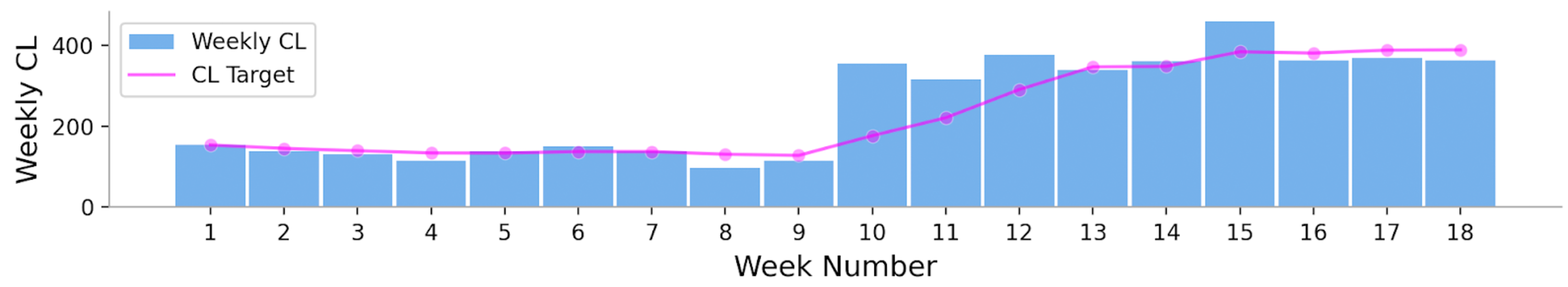}
\caption{Weekly Cardio Load and target with a step increase in weekly Cardio Load at week 10.}
\label{fig:target_adaptation_2}
\end{figure}

\begin{figure}[hbt]
\centering
\includegraphics[width=.8\linewidth]{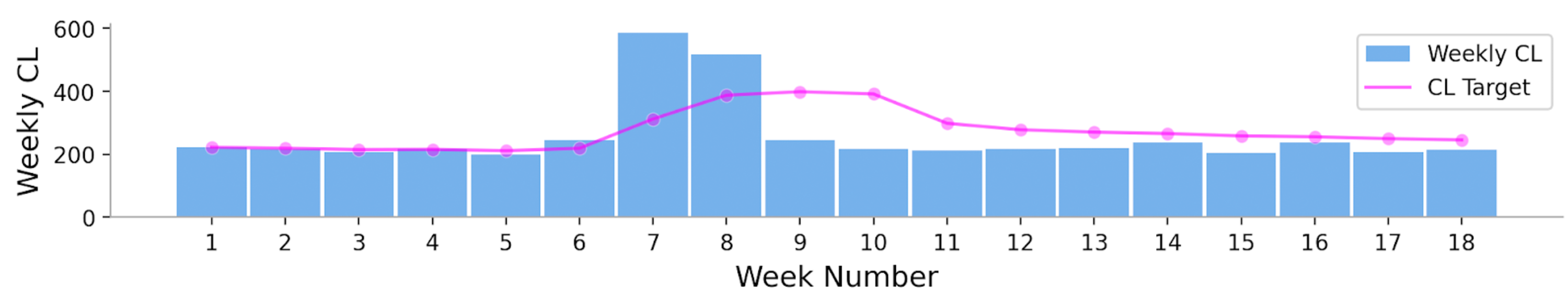}
\caption{Weekly Cardio Load and target with an acute increase in weekly Cardio Load in weeks 7-8.}
\label{fig:target_adaptation_3}
\end{figure}

\cref{fig:target_adaptation_2,fig:target_adaptation_3} show examples of two different users' weekly Cardio Load showing just CL and CL Target over the same period. 
The graphs also show how the calculated target loads adapt to changes in the weekly load. 

In \cref{fig:target_adaptation_2} the weekly Cardio Load is steady until week 10 when the user increases their weekly load to a consistently higher level. 
The response is a steady upward increase of the target load as it adapts to the user's new routine. The target load does not rise too quickly to avoid the risk of over-training. 
Once a steady state is reached then the new target guides the user towards maintaining their higher fitness level.

\cref{fig:target_adaptation_3} shows a case where there is a rapid increase in weekly cardio load, however after two weeks the load reverts back to the previous level. Again the target load increases, but following the drop in Cardio Load it remains elevated for a period of time to give the user the opportunity to build on their success and try to maintain their new level. As the Cardio Load remains low, the target slowly returns back to its previous baseline level.

In these examples the adaptation of target load takes place in a way that motivates the user to maintain (or exceed) their fitness level without the risk of excessive overtraining, while not setting an unrealistically high, potentially demotivating target.

\subsection{Minimum target load}

For all users, when the weekly CL is reduced for a significant time, for example during recovery from illness or injury, the CL target may fall significantly from its usual range. To avoid setting a very low target, which could be unhelpful and demotivating, we will specify a minimum target load. Since Active Zone Minutes (AZMs) are aligned with health guidelines, minimum target loads were calculated by comparing Cardio Load and AZM distributions in a population of ten thousand consenting Fitbit users. At present a single value is applied for all users, however in future targets adjusted for age group and gender may be applied. This approach ensures that users meeting the weekly target load will meet the lower thresholds of health guidelines.

\subsection{Target load during onboarding, wear requirements and missing data }

When a user signs into the Fitbit app for the first time, their initial target load is the minimum target load, calculated as described in the previous section. For the first few weeks, users should try to wear the watch every day for a substantial part of the day. The first personalized target load appears after 7 days of data have been acquired, then the value is refined in subsequent weeks. After 4 weeks, the best personalized estimate possible from the data is presented.  

Users should try to wear their compatible Fitbit tracker or Pixel watch every day. Where possible the device should be worn during workouts and as much as possible during the day to capture activities outside of structured workouts.

Non-wear will inevitably result in missing opportunities to record activity and as a result the weekly Cardio Load will be reduced. For example, not wearing the device for a day could cause the Cardio Load to be significantly lower than if the user had worn the device, whether they complete workouts or not. As the target load is calculated weekly, then missing a day would not severely impact the target load as it is influenced by the previous three weeks data (provided this is available). Consistently missing segments of the day, especially when the user is active will also negatively impact the target load.

\section{Summary}
In summary\footnote{The content, features, and user interface elements shown in this research paper are conceptual and subject to change. The final product may differ significantly. This research paper is a preview of anticipated functionality and is for illustrative purposes only. We reserve the right to modify or remove any features prior to launch.}, Google Fitbit's Cardio Load, launched with Pixel Watch 3 and refined in the Public Preview of the new Fitbit app is a personalized weekly metric capturing cardiovascular load from all activity intensities. It's based on heart rate reserve and exponentially weights higher intensities.

Key features and benefits:
\begin{itemize}
\item
{\bf Comprehensive Measurement:} Cardio Load considers both structured workouts and incidental daily activities, offering a holistic view of exertion.
\item
{\bf Distinction from AZMs:} Unlike Active Zone Minutes (AZMs) which focus on general health guidelines, Cardio Load is designed for performance measurement and improvement, rewarding incremental changes in intensity.
\item
{\bf Adaptive Weekly Targets:} Personalized weekly targets are set based on a rolling average of past Cardio Load, adapting to individual training patterns and encouraging consistent progress without risk of overtraining.
\item
{\bf Motivational Tool:} The adaptive targets and visual progress tracking in the Fitbit app help users plan workouts, catch up on load, and stay motivated towards their fitness goals.
\item
{\bf Minimum Target Load:} A minimum target load, adjusted for age and gender and aligned with health guidelines, ensures users meet a baseline level of activity.
\end{itemize}

\bibliography{main}

\end{document}